\documentclass[twocolumn]{article}
\usepackage{amsmath,amssymb,amsfonts}
\usepackage{algorithmic}
\usepackage{optidef}
\usepackage{graphicx}
\usepackage{textcomp}
\usepackage{xcolor, colortbl}
\usepackage{tikz}
\usepackage{booktabs}

\usepackage[sorting=none]{biblatex} 
\newtheorem{property}{Property}

\newcommand\keywords[1]%
  {\begin{flushleft}
   \let\and\\%
   \textbf{Keywords:}\\
   #1
   \end{flushleft}%
  }

\begin{document}

\title{Neural-powered unit disk graph embedding: qubits connectivity for some QUBO problems}


\DeclareRobustCommand*{\IEEEauthorrefmark}[1]{%
  \raisebox{0pt}[0pt][0pt]{\textsuperscript{\footnotesize #1}}%
}

\author{
Chiara Vercellino\IEEEauthorrefmark{1}\IEEEauthorrefmark{*}
\and  Paolo Viviani\IEEEauthorrefmark{1}
\and Giacomo Vitali\IEEEauthorrefmark{1}
\and Alberto Scionti\IEEEauthorrefmark{1}
\and Andrea Scarabosio\IEEEauthorrefmark{1}
\and Olivier Terzo\IEEEauthorrefmark{1}
\and Edoardo Giusto\IEEEauthorrefmark{2}
\and Bartolomeo Montrucchio\IEEEauthorrefmark{2}\vspace{0.5cm}\\ 
\IEEEauthorrefmark{1}\textit{LINKS Foundation}, Torino, Italy \\
\IEEEauthorrefmark{2}\textit{\textit{DAUIN}, Politecnico di Torino}, Torino, Italy\\
\IEEEauthorrefmark{*}\textit{chiara.vercellino@linksfoundation.com}
}

\maketitle
\begin{abstract}
Graph embedding is a recurrent problem in quantum computing, for instance, quantum annealers need to solve a minor graph embedding in order to map a given Quadratic Unconstrained Binary Optimization (QUBO) problem onto their internal connectivity pattern. This work presents a novel approach to constrained \emph{unit disk} graph embedding, which is encountered when trying to solve combinatorial optimization problems in QUBO form, using quantum hardware based on neutral Rydberg atoms. The qubits, physically represented by the atoms, are excited to the Rydberg state through laser pulses. Whenever qubits pairs are closer together than the blockade radius, entanglement can be reached, thus preventing entangled qubits to be simultaneously in the excited state. Hence, the blockade radius determines the adjacency pattern among qubits, corresponding to a unit disk configuration. Although it is straightforward to compute the adjacency pattern given the qubits' coordinates, identifying a feasible unit disk arrangement that matches the desired QUBO matrix is, on the other hand, a much harder task.
In the context of quantum optimization, this issue translates into the physical placement of the qubits in the 2D/3D register to match the machine's Ising-like Hamiltonian with the QUBO formulation of the optimization problems. The proposed solution exploits the power of neural networks to transform an initial embedding configuration, which does not match the quantum hardware requirements or does not account for the unit disk property, into a feasible embedding properly representing the target optimization problems. Experimental results show that this new approach overcomes in performance \textit{Gurobi} solver.
\end{abstract}




\section{Introduction}\label{sec:intro}
Quantum computers bear promises as tools to accelerate specific computations like prime factorization, eigenvalues decomposition, combinatorial problem-solving etc. However, in the noisy intermediate-scale quantum (NISQ) era, practical applications of canonical quantum algorithms (\textit{e.g.}, Shor's algorithm) are still out of reach due to low qubits count, limited coherence time, and gates fidelity; in this context different approaches emerged, like hybrid quantum-classical algorithms and specialised architectures (\textit{e.g.}, \emph{quantum annealers, quantum simulators}), that showed promising results for specific problems, but still present a number of implementation issues.

The goal of this work is to provide a suitable methodology to port real-life applications to these kinds of machines, with a specific focus on solving discrete optimization problems on optical quantum simulators based on neutral atoms, by developing a Machine Learning (ML) technique for embedding graphs into the suitable machine representation (register).

This quantum computing architecture operates at room temperature and it is, in principle, flexible enough to implement quantum gates as well as manipulate a given target system Hamiltonian through laser pulses \cite{henrietQuantumComputingNeutral2020}. In this paper, gates-based operations will not be considered, instead, attention will be paid to the analog quantum processing mode, which involves the use of laser pulses to induce dipole-dipole interactions between Rubidium atoms (\textit{i.e.}, Rydberg atoms \cite{gallagher1988rydberg}) in \emph{Rydberg state}, arranged arbitrarily in 2D or 3D arrays.

The interactions between Rydberg atoms can, in turn, be mapped into a spin Hamiltonian (\textit{e.g.}, Ising): this capability of neutral atoms machines is interesting because a significant set of NP-hard combinatorial optimization problems can be mapped to the Ising Hamiltonian \cite{lucasIsingFormulationsMany2014} or to the equivalent Quadratic Unconstrained Binary Optimization (QUBO) forms \cite{glover2018tutorial}, popularized by Quantum Annealers.
In principle, once a set of Rydberg atoms is organised in the space to reproduce the desired Hamiltonian, it is possible to solve the associated optimization problem using the Quantum Approximate Optimization Algorithm \cite{serretSolvingOptimizationProblems2020}, \cite{pasqal_optim}.

While the idea is somewhat straightforward, the hardware requirements of the experimental machine characterize the optimization problems that can be approached.
The starting point is to model the optimization problem at hand in the QUBO form, which means building a square matrix $Q$ $\in \mathbb{R}^{n \times n}$ whose dimensionality $n$ is determined by the number of binary variables.
In the proposed embedding solution, each binary variable is represented by a physical qubit.
The diagonal elements of $Q$ identify the penalty/gain associated with each binary variable when it assumes the value $1$; due to the restriction of the machine lasers to operate only at the global level (\textit{i.e}., exciting all atoms with the same Rabi frequency \cite{henrietQuantumComputingNeutral2020}), these diagonal elements should have all the same value.
The off-diagonal elements of $Q$ represent quadratic interactions between the variables, which are represented by atoms coupled through the Rydberg blockade effect \cite{urban2009observation} which scales as $\propto 1/{d_{ij}^6}$, with $d_{ij}$ the Euclidean distance between two qubits $i$ and $j$.
Thus, as the distance is necessarily positive, the off-diagonal elements of $Q$ should have all the same sign.

\begin{figure}[t]
\centerline{\includegraphics[width=1.0\columnwidth]{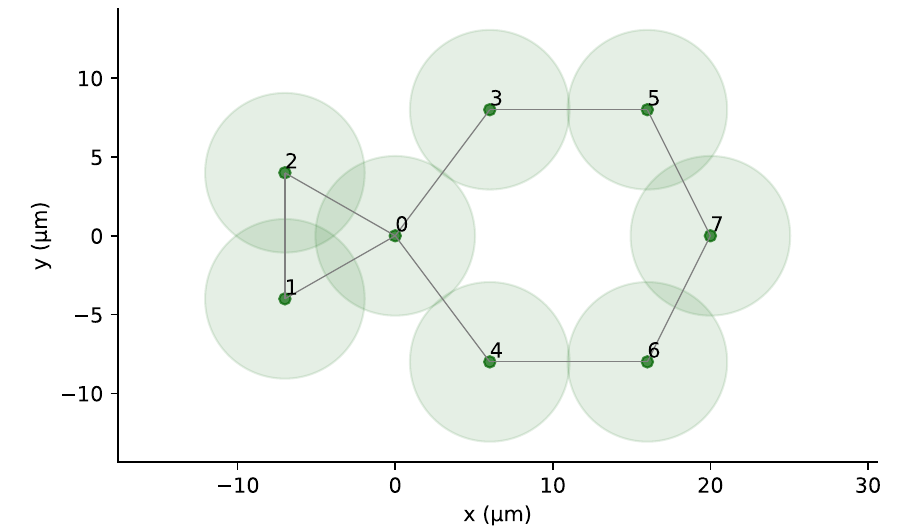}}
\caption{Representation of a qubit register composed of neutral atoms arranged in a 2D array.
Interaction between qubits is regulated by the Rydberg blockade effect: only atoms falling within the Rydberg radius are subject to interaction.}
\label{ud_graph}
\end{figure}

If a QUBO problem satisfies the requirements above, it is in principle straightforwardly mapped into the effective Hamiltonian that the real machine is able to reproduce
\begin{equation}\label{hamiltonian_eq}
    \mathcal{H} = \sum_{i=1}^{n} \frac{\hslash \Omega}{2} \sigma_{i}^{x} - \sum_{i=1}^{n} \frac{\hslash \delta}{2} \sigma_{i}^{z} + \sum_{j>i} \frac{C_6}{d_{ij}^{6}} n_i n_j
\end{equation}
where $n_i = (1+\sigma_i^z)/2$ is the Rydberg state occupancy, $\sigma^{x,z}_i$ are the Pauli matrices of the spin of $i$-th qubit, $\Omega$ is the Rabi frequency of the laser pulse and $\delta$ is the detuning of the laser pulse.
While a feasible formulation of the QUBO matrix based on these constraints can be achieved (\textit{e.g.}, for graph coloring problems \cite{glover2018tutorial}), the magnitude of the off-diagonal elements of $Q$ is bound to the two-qubits coupling coefficient $C_6/{d_{ij}^6}$, therefore to the relative distances of qubits in a 2D/3D register.
In particular, a threshold-like effect is reached when the distances between qubits' pairs are shorter than the blockade radius, \textit{i.e.}, a critical distance at which the strength of the interactions balances with the Rabi frequency, thus yielding opposite entanglement \cite{PhysRevLett.104.010502} on neutral atoms: atoms with a pair distance shorter than the blockade radius cannot be simultaneously in the excited state.
Therefore, the interactions between qubits in the register can be represented by a unit disk (UD) graph \cite{CLARK1990165} as in figure \ref{ud_graph}, and the problem of mapping an arbitrary QUBO/Ising problem onto a neutral atoms quantum simulator can be formulated as a unit disk graph embedding problem. 

This paper will describe a novel heuristic to obtain the coordinates of a unit disk embedding for a given QUBO problem.
\section{Related Work}
The qubits placement problem is well-known in the context of quantum computing. Despite the specific technologies, the design of a proper embedding to exploit the hardware potential is of major importance. In the particular case of neutral atoms based devices, the embedding for QUBO problems has been approached in different ways.

In \cite{pasqal_optim}, non-convex constrained formulations of the UD graph problem are provided, anyway, solving the UD graph problem with classical solvers can become pretty difficult as the size of the QUBO problems increases or the adjacency pattern gets more complicated: the computation of the unit disk graph could become possibly harder than retrieving the QUBO problem solution. To fully understand the complexity of the unit disk graph problem, it is important to notice that even the corresponding recognition problem (\textit{i.e.}, determining if a given graph has a realization that maps vertices to points of a unit disk configuration) is NP-hard \cite{breuUnitDiskGraph1998}.\\ Further results on UD graphs are presented in \cite{kuhnUnitDiskGraph2004a} with a focus on the non-approximability of the unit disk (and d-quasi unit disk \cite{kuhn2003ad}) graphs, moreover, a bound on the quality of the embedding is computed. In \cite{bhoreUnitDiskRepresentations2021}, S. Bhore et al. investigate the unit disk graph recognition problem for subclasses of planar graphs, stating that even for outerplanar \cite{syslo1979characterizations} and trees \cite{west2001introduction} graphs this task is NP-hard.

Given the complexity of the problem, to overcome the limitations of an exact solution, embedding heuristics have been proposed, but their application to the neutral atoms quantum hardware generally needs more physical qubits or greater sophistication in the hardware design. Several studies approach the embedding issue through local fields' manipulation. In \cite{lechner2015quantum}, local interactions enhance the quantum annealing architecture with controllable all-to-all connectivity between logical qubits, the physical qubits are instead placed according to a square-lattice geometry.
In \cite{qiu2020programmable}, the authors illustrate a three-dimensional embedding based on Ising ferromagnetic quantum wires to couple distant qubits and produce a regular cubic lattice architecture. As additional requirements to build quantum wires, multiple physical qubits are used to represent a single variable and local fields play a key role. Further work on quantum wires is presented in \cite{kimRydbergQuantumWires2021}, where Kim et al. overcome the issue of local fields in quantum wires for maximum independent set problems, anyway, the proposed architecture still requires auxiliary wire atoms in a 3D space and constrains the embedding to accomplish constant distances among adjacent qubits, thus introducing a discretization of the embedding space and not exploiting the full potential of the blockade effect.

QUBO problems' embedding has also been approached through graph minors \cite{robertson1995graph} search. The D-Wave system exploits heuristic techniques to find graph minors with a probabilistic setting to avoid exhaustive search \cite{caiPracticalHeuristicFinding2014}. Their quantum hardware is designed with a fixed topology where qubits adjacency is modelled through couplers. The mapping of the QUBO problems is feasible whenever the graph of binary optimization variables interactions is a
minor of the graph of qubit interactions in the D-Wave hardware. A similar approach was presented also by V. Choi \cite{choi2008minor}, with a focus on embedding Ising Hamiltonian for solving the maximum independent set problem via adiabatic quantum computation. Again the reduction through minor-embedding in the quantum hardware graph relies on parameters settings of qubit biases and coupler strengths.

\section{Methodology}
Starting from a properly representable, \textit{i.e.}, compliant to the Hamiltonian formulation of eq. \eqref{hamiltonian_eq}, QUBO model, a feasible embedding for the qubits should be retrieved. The adjacency matrix $A$, associated with matrix $Q$, defines the adjacency pattern among the qubits, but the concept of proximity/distance of the atoms in the register cannot be defined at will. 
Qubits placement has to satisfy strong requirements in order to reproduce the blockade effect that underlies the entanglement between them and, therefore to enhance optimization problems solution with quantum technology.

To model these embedding constraints, the following notation will be used: $\mathcal{G}=(\mathcal{V}, \mathcal{E})$ is the graph associated with matrix $A$, with $\mathcal{V}$ vertexes and $\mathcal{E}$ edges, $\mathcal{G'}=(\mathcal{V}, \mathcal{\bar{E}})$ is the complement graph of  $\mathcal{G}$. Finally, each vertex in $\mathcal{V}$ represents a qubit and $x_i$, $y_i$ and possibly $z_i$ are the coordinates of the qubits into the 2D/3D domain, $\forall i \in \mathcal{V}$.

\subsection{From QPU requirements to embedding constraints}
The register on which qubits have to be placed identifies the domain for the qubits' positions. In the considered device (Pasqal's R\&D prototype, Chadoq2\cite{silverioPulserOpensourcePackage2022}), this domain can cover at most a circular area with a maximum radius of $50\ \mu m$, so qubits must be placed with a maximum pair distance $d_{ij}$ of $100\ \mu m$. Moreover, in this machine setting, two atoms cannot be closer than $4\ \mu m$.
Hence, the constraints: 
\begin{align}
    d_{ij} &\leq 100\ \mu m &\qquad \forall i,j \in \mathcal{V},\ i < j \label{eq:max} \\
    d_{ij} &\geq 4\ \mu m &\qquad \forall i,j \in \mathcal{V},\ i < j \label{eq:min}
\end{align}
Then, the unit disk representation requires vertexes, \textit{i.e.}, qubits, that share an edge, to be closer than non-adjacent vertexes, \textit{i.e.}, the ones not connected by an edge. In principle, this constraint could be written as:
\begin{equation}\label{eq:ud_constr}
    \max_{(i,j) \in \mathcal{E}}{d_{ij}} < \min_{(h,k) \in \mathcal{\bar{E}}}{d_{hk}}
\end{equation}
But this formulation could lead to an embedding configuration that prevents the blockade effect to take place.
In fact, the Rydberg blockade effect occurs in the regime of strong interactions, which determines the blockade radius $r_b$ \cite{ciampini2015ultracold} as 
\begin{equation}\label{eq:rabi_from_freq}
    r_b = \sqrt[6]{ \frac{C_6}{\hslash \Omega}}
\end{equation}
Moreover, to obtain the opposite entanglement \cite{picken2018entanglement} between pairs of qubits that fall into the $r_b$ radius, the laser pulses should have a time duration that is \begin{equation}\label{eq:time_from_freq}
    t = \frac{\pi}{\sqrt{2} \Omega}
\end{equation}
This is the time duration at which the maximally entangled state is reached \cite{PhysRevLett.104.010502}.
From equation~\eqref{eq:time_from_freq}, it is possible to compute the minimum Rabi frequency that can be allowed given $\Tilde{t}$, the coherence time limit of the machine.
Then, exploiting eq.~\eqref{eq:rabi_from_freq} the maximum blockade radius $\Tilde{r_b}$ can be obtained as follows:
\begin{equation}
    \Tilde{r_b} = \sqrt[6]{\frac{\sqrt{2}|C_6|\Tilde{t}}{\hslash \pi}}
\end{equation}

This leads to a reformulation of constraint~\eqref{eq:ud_constr} that considers also the time limit $\Tilde{t}$ of the machine:
\begin{align}
    d_{ij} \leq \Tilde{r_b} &\qquad \forall (i,j) \in \mathcal{E} \label{eq:adj} \\ 
    d_{hk} > \Tilde{r_b} &\qquad \forall (h,k) \in \mathcal{\bar{E}} \label{eq:not_adj}
\end{align}

The described constraints define a constrained UD graph problem, whose solution, both in the 2D and 3D cases, provides realistic coordinates for qubits embedding in the real quantum register. Providing qubits’ positions, while avoiding fixed geometries, is enhanced by the capability to control individual atoms trapped in optical tweezers, through a real-time control system \cite{barredo2016atom}.

\subsection{Traditional approaches as a starting point}
Starting from the above-mentioned constraints, the first approach to the UD embedding problem could be to solve the corresponding quadratic constrained non-convex optimization problem that arises using the positions of the qubits as variables. For these kinds of problems, several solvers have been proposed (\textit{e.g.}, \cite{park2017general, doi:10.1080/10556788.2017.1350675, hart2011pyomo, bynum2021pyomo}), but the UD embedding problem in the general case is NP-hard, so these approaches did not provide feasible solutions as the number of qubits increases or as the connectivity pattern becomes more complicated.

Other approaches to achieving embedding representation take inspiration from force-directed algorithms, like the Fruchterman-Reingold method \cite{fruchterman1991graph} \cite{hagberg2008exploring}: this force-directed layout algorithm models attractive and repulsive forces between vertexes pairs, according to the adjacency pattern described by $A$; repulsive forces intervene on all vertexes pairs with module $k^2/d_{ij}^2$, whilst attractive forces intervene only on adjacent pairs and have module $d_{ij}/k$. The traditional algorithm, in this case, is designed to find the embedding in a square domain and takes as a parameter the optimal distance $k$ at which the two forces balance for adjacent pairs \cite{hagberg2008exploring}, which in our case could be set to a value in the range $[4, \Tilde{r_b}]$. There exists an easy way to modify the domain of the embedding to match our requirement expressed by eq. \eqref{eq:max}, that is by just changing the positions' projector into a circular domain with a radius of $50\ \mu m$ instead of the current square domain projector, anyway, a further modification to match all other constraints (eqs.~\eqref{eq:min},~\eqref{eq:adj},~\eqref{eq:not_adj}) is not straightforward.

Finally, it is worth mentioning that some optimization problems inherit their connectivity patterns in the QUBO formulation from their own topology, such as antennas' positions in a PCI (Physical Cell Identifier) problem \cite{Kavlak2012}, \cite{Gui2019}. In these cases, some initial positions are yet provided and they can be scaled to match the register limit imposed by eq.~\eqref{eq:max}, anyway the matching of all other constraints is not guaranteed.\\
All these approaches provide some initial positions, $(x_i, y_i)\ \forall i \in \mathcal{V}$, for the UD embedding in the register, which is yet an improvement over a randomly set embedding, nevertheless, further manipulation is required to match the optimization problem structure.

\subsection{GEAN: Graph Embedding Autoencoder Network}
\subsubsection{Model architecture}
\begin{figure*}[ht!] 
    \centerline{
    \includegraphics[width=1.0\linewidth]{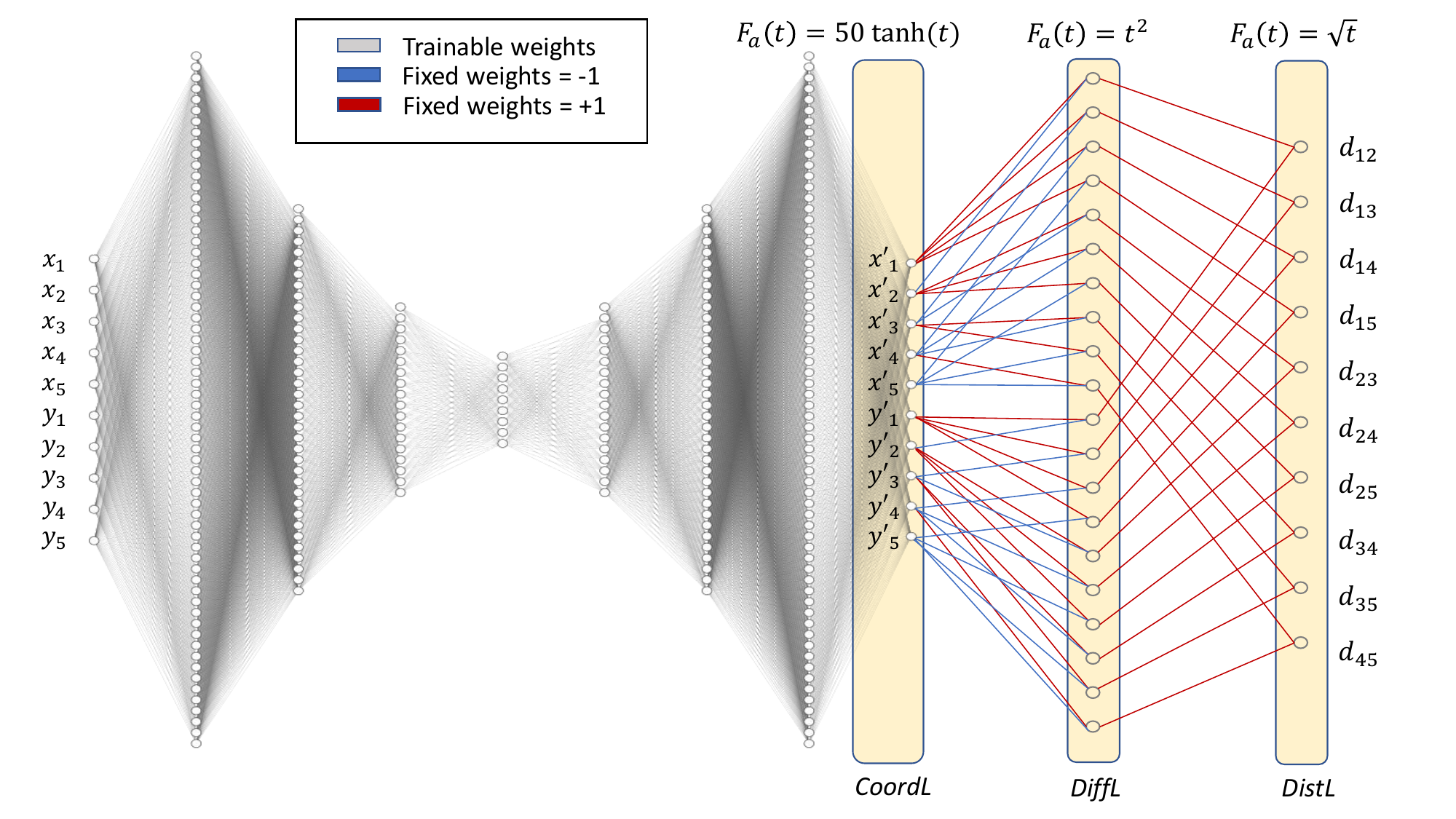}}
    \caption{GEAN architecture: 5 qubits embedding in 2D, the initial coordinates $(x_1,y_1),\ldots (x_5,y_5)$ are transformed into feasible coordinates $(x^{'}_{1},y^{'}_{1}),\ldots (x^{'}_{5},y^{'}_{5})$. The output layer of the model, \textit{DistL}, provides all pairs' distances between nodes.}
    \label{fig:nn_architecture}
\end{figure*}

The proposed solution implements a Neural Network based model, reported in figure~\ref{fig:nn_architecture}, which has two main components:

\textit{i)} The autoencoder part follows the typical symmetric architecture, with an input layer whose nodes are associated with the initial positions $(x_i, y_i) \ \forall i\in\mathcal{V}$ and an output layer, \textit{CoordL}, that provides the transformed positions $(x^{'}_i, y^{'}_i)\ \forall i\in\mathcal{V}$.
The \textit{CoordL} is equipped with an activation function that is defined as $F_a(t):=50tanh(t)$, thus the transformed coordinates belong to the square domain $(-50,50)^2$ that inscribes the circular domain of interest.
As the considered domain is two-dimensional, the input and the output layers have $2n$ nodes each, the first $n$ nodes of each layer represent x-coordinates, and the other $n$ nodes are y-coordinates.
The hidden layers of the model respectively have $64, 36, 18, 9, 18, 36$ and $64$ hidden nodes, they use ReLU as activation functions, and they are interspersed with dropout layers, with dropout probability set to $0.5$, for regularization purposes.
All the weights of the autoencoder are trainable and the bias nodes' contributions are also taken into account. The choice of this particular architecture was driven by two main observations. On one hand, as the order of the elements in the input vector does not imply local correlations, Convolutional Neural Networks (CNNs) have been discarded, as fully-connected architecture allowed a more general interpretation of the elements in the input vector, we opted for that choice; on the other hand, autoencoder architecture was proposed in \cite{eguchi2022ig}, in this work, an autoencoder is used to retrieve coordinates given the distance matrix in a protein modelling use case, even though the same architecture was not straightly usable for our use case, we do not have feasible pair distances a priori, Eguchi et al. study suggested us that our neural network architecture could benefit from a latent embedding representation.

\textit{ii)} The second component makes the model aware that the output of the autoencoder, $(x^{'}_i, y^{'}_i)\ \forall i\in\mathcal{V}$, are Cartesian coordinates.
This component is in charge of computing Euclidean distances between all the qubits' pairs, which is obtained by adding a sparse layer that connects the \textit{CoordL} to nodes in the \textit{DiffL} layer, which consists of $2 \binom{n}{2}=n(n-1)$ nodes. The \textit{DiffL} layer has no contribution from a bias node and non-trainable weights in the sparse layer are set to $\pm 1$, such that the nodes $\alpha$ in the \textit{DiffL} reflect differences between pairs of coordinates:
\begin{align}
    \alpha_{(i-1)(n-1)-\binom{i-1}{2}+j-i} &= x^{'}_i-x^{'}_j \quad i,j \in \mathcal{V},\ i<j \\
    \alpha_{(n-1)(\frac{n}{2} + i-1)-\binom{i-1}{2}+j-i} &= y^{'}_i-y^{'}_j \quad i,j \in \mathcal{V},\ i<j
\end{align}
Then a square activation function $F_a(t):= t^2$ is applied to \textit{DiffL}. Finally, the squared pairs' distances are retrieved by adding another sparse layer, with fixed weights of value $+1$ that allows the $\binom{n}{2}$ nodes, $\beta$, in the output layer, \textit{DistL}, to assume the following values:
\begin{equation}
    \beta_k = \alpha^{2}_k + \alpha^{2}_{\binom{n}{2}+k} \qquad k \in \Big\{1,2,\ldots, \binom{n}{2}\Big\}
\end{equation}
At last, the $F_a(t):=\sqrt{t}$ activation function intervenes on the \textit{DistL} layer to make the outputs of the GEAN model, $d$, become the pairs' distances $d_{ij},\ i<j \in \mathcal{V}$.

\subsubsection{Loss function}
Once the pairs' distances are computed, it is possible to model the loss function to target a feasible UD embedding. In this case, the training of the model has a different meaning from the one of a typical machine learning task: the mini-batch of the training set consists of just one sample, \textit{i.e.}, the initial positions $(x_i, y_i) \ \forall i\in\mathcal{V}$, and the minimization of the loss function does not correspond to the intent of prediction, but it pursues a proper approximation of the domain transformation to achieve the desired UD connectivity.\\
As there are four kinds of constraints that the UD embedding must satisfy (eqs.~\eqref{eq:max},~\eqref{eq:min},~\eqref{eq:adj},~\eqref{eq:not_adj}), the proposed loss function is four-folded and each part of it is applied to the distances output vector $d$: in accordance to eq.~\eqref{eq:max}, the qubits' pairs that are more distant than $100\ \mu m$ should be penalized, so the first contribution to the loss function, $loss_1$, is computed as follows:
\begin{equation}
    loss_1(d) = \underset{i<j \in \mathcal{V}}{avg} |max(100, d_{ij})-100|
\end{equation}
To match constraint~\eqref{eq:min}, which defines the minimum distance allowed between qubits' pairs, $loss_2$ contribution is designed as:
\begin{equation}
    loss_2(d) = \underset{i<j \in \mathcal{V}}{avg} |min(4, d_{ij})-4|
\end{equation}
Then, $loss_3$ expresses constraint~\eqref{eq:adj}, and it penalizes pairs' distances of adjacent qubits that are greater than $\Tilde{r_b}$:
\begin{equation}
    loss_3(d) =\underset{i<j \in \mathcal{V}}{avg} A_{ij}|max(\Tilde{r_b}, d_{ij})-\Tilde{r_b}|
\end{equation}
Finally, $loss_4$ is the counterpart for constraint~\eqref{eq:not_adj}, and it penalizes pairs' distances of non-adjacent qubits that are shorter than $\Tilde{r_b}+\epsilon$, where parameter $\epsilon$ is set to $0.1$ to account for the strictly greater sign in the constraint:
\begin{equation}
    loss_4(d) =\underset{i<j \in \mathcal{V}}{avg} (1-A_{ij}) |min(\Tilde{r_b}+\epsilon, d_{ij})-(\Tilde{r_b}+\epsilon)|
\end{equation}

The overall loss function sums up all the previous contributions. The absence of weighting constants in the final loss formulation is justified by the fact that all constraints violations have the same importance; they all would imply an unfeasible embedding:
\begin{equation}
    loss(d)=loss_1(d)+loss_2(d)+loss_3(d)+loss_4(d)
\end{equation}
In order to minimize the $loss$ function, the \textit{PyTorch} implementation  of the AdamW algorithm \cite{loshchilov2017decoupled} has been exploited, in combination with a constant learning rate equal to $1e^{-3}$. The maximum number of epochs allowed for each graph embedding has been set to $5000$, but a stopping criterion based on the achievement of a feasible embedding prevents the training to go through all the epochs when it is not necessary.

\subsection{The limitations of the 2D domain}
\begin{figure}[ht!]
\centerline{\includegraphics[width=1.0\columnwidth]{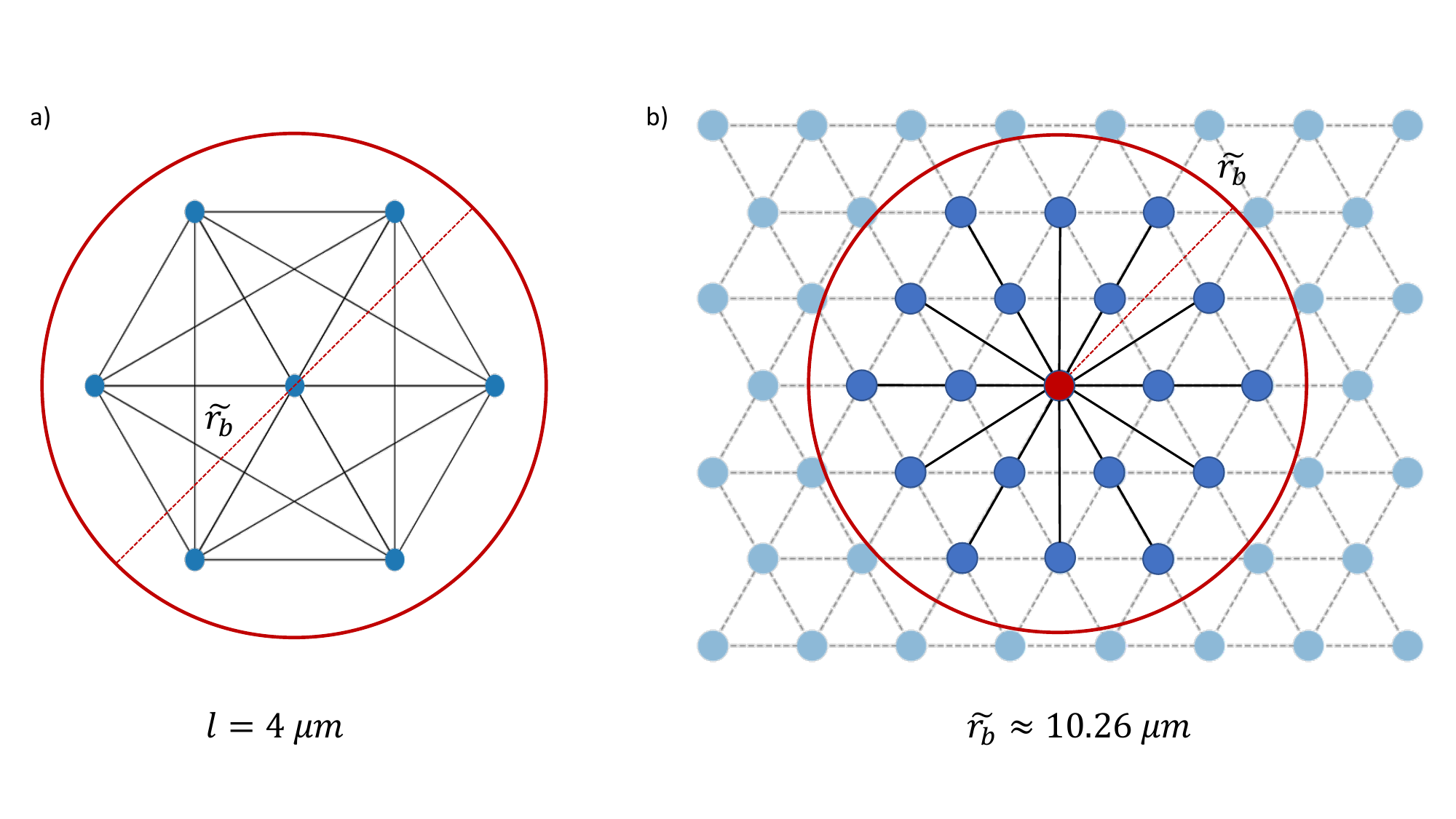}}
\caption{Representation of the maximal clique, $K_7$, fig. \ref{fig:triangular_embedding}a), and of the maximal degree, fig. \ref{fig:triangular_embedding}b), that can be embedded on the register. The baseline embedding is the regular hexagonal packing with $l=4\ \mu m$ side and the Rydberg blockade radius is the maximum available on the machine $\Tilde{r_b} \approx 10.26\ \mu m$.}
\label{fig:triangular_embedding}
\end{figure}

This UD embedding methodology has to be subject to the limitations of 2-dimensional space, thus the connectivity of the graphs that can be embedded is not general. As they are of particular interest to identify which QUBO problems could be embedded into the 2D register, two necessary conditions have been identified: one concerning the maximal clique and the other one regarding the maximal degree of graph $\mathcal{G}$. These two connectivity-related properties should be complemented by the other requirements expressed in the \ref{sec:intro} section: constant values in the diagonal of $Q$ and constant sign for its off-diagonal elements.
The key concept underlying these results comes from \emph{Thue's Theorem}, which states that \textit{the regular hexagonal packing is the densest circle packing in the plane} \cite{chang2010simple}. Thus, the densest packing for the considered qubits' register has the structure reported in figure~\ref{fig:triangular_embedding}b). Considering constraint~\eqref{eq:min}, the hexagon side of the densest embedding is $4\ \mu m$. Furthermore, the time limit of the machine determines the maximum $\Tilde{t}$, and consequently $\Tilde{r_b}$,  for this experimental setting $\Tilde{t}=3\ \mu s$ and $\Tilde{r_b} \approx 10.26\ \mu m$. From these considerations, the following properties are obtained:

\begin{property}[Maximum clique property]
The maximum $N$, such that a complete graph (or clique), $K_N$, with $N$ vertices, can be embedded into the register is $N=7$.
\end{property} 

\begin{property}[Maximum degree property]
Given a graph $\mathcal{G}(\mathcal{V},\mathcal{E})$, the maximal degree $d_{max}(\mathcal{G})$ allowed for a feasible embedding is $18$.
\end{property} 

Fig.~\ref{fig:triangular_embedding} represents both these properties. To embed a clique, all the vertices of the graph, $K_N$, must lie within a circle of maximum radius set to $\frac{\Tilde{r_b}}{2}$~(fig. \ref{fig:triangular_embedding}a)) and the maximum degree for a graph, $d_{max}(\mathcal{G})$, implies all the adjacent qubits to lie within a maximum radius $\Tilde{r_b}$- Anyway this bound on the maximum degree relies on the hypothesis that the adjacency pattern among all other qubits within the circle matches the adjacency matrix $A$ one.

\subsection{Adding new dimensions: the 3D embedding case}

To be able to embed a greater set of QUBO problems and with a view to a future quantum machine able to work with a 3D register, the GEAN model can be modified to obtain also feasible z-coordinates. This implies some modification to the architecture presented in Fig.~\ref{fig:nn_architecture} but does not require loss function changes, given that the constraints on pairs' distances remain the same. The architecture changes concern:

\textit{i)} the addition of nodes in the input and in the \textit{CoordL} layers to represent the z-coordinates, $z_i$ and $z^{'}_i$, $\forall i \in \mathcal{V}$, so the size of these layers passes from $2n$ to $3n$ nodes. When the initial $z_i$ coordinates are not available, they can be initialized to $0$;

\textit{ii)} the \textit{DiffL} layer should be modified to account also for the z-coordinates pairs' differences, so it contains $3 \binom{n}{2}$ nodes, and the weights in the fully connected layer are modified such that the last $\binom{n}{2}$ $\alpha$ nodes in \textit{DiffL} are
\begin{equation}
    \alpha_{(n-1)(n + i-1)-\binom{i-1}{2}+j-i}=z^{'}_i-z^{'}_j \quad i,j \in \mathcal{V},\ i<j
\end{equation}

\textit{iii)} the \textit{DistL} layer does not change the size, as it still computes pairs' distances between qubits, but the increment of the \textit{DiffL} layer requires it to consider the contribution also from the z-coordinates square differences, so $\beta$ nodes input values become
\begin{equation}
    \beta_k = \alpha^{2}_k + \alpha^{2}_{\binom{n}{2}+k} + \alpha^{2}_{2\binom{n}{2}+k} \qquad k \in \Big\{1,2,\ldots, \binom{n}{2}\Big\}
\end{equation}
\section{Results and Discussion}

\subsection{Classical solver as a benchmark}

In order to provide benchmarks, we formulated the Unit Disk graph problem and attempted to find solutions for feasible embeddings with state-of-the-art solvers. In particular, we opted for the \textit{Gurobi} solver \cite{bixby2007gurobi} and handled the implementation of the programming model through the \textit{Pyomo}\footnote{https://pyomo.readthedocs.io} Python library. The choice of a classical solver for comparison, instead of a heuristic approach based on fixed geometries, is intended to try and fully exploit at most all the degrees of freedom available for the placements of the atoms in the register.\\ Since the considered optimization problem is NP-hard, the computing time for retrieving the optimal solution is linked to the dimension of the problem, \textit{i.e.}, the $n$ number of qubits. Accordingly, as $n$ increases it becomes necessary to set up a time limit for the solver, thus we set a maximum walltime for the classical solver to $2$ minutes, as it is the maximum time required by the GEAN model to retrieve a feasible solution on the largest instance.\\ In the definition of the programming model, which is quadratic and non-convex, we will refer to $\mathcal{P}$ as the set of all unordered pairs in $\mathcal{V}$, thus $|\mathcal{P}|=\frac{n(n-1)}{2}$, the embedding dimensionality is $N \in \{2,3\}$, and the positions of the qubits will be represented by $N$-dimensional vector $\overrightarrow{p}_i$, $\forall i \in \mathcal{V}$. The retrieval of a feasible embedding is modelled through $\gamma_{ij}$ binary variables, described as follows.

\begin{equation}
\gamma_{ij} := 
\begin{cases}
	1 & $pair distance is unfeasible $\\
	0 & $pair distance is feasible $
\end{cases}
\end{equation}

Hereafter, we present the overall constrained Unit Disk graph problem, with reference to the specifications of the considered Quantum machine.

\begin{mini}|s|
{\overrightarrow{p}, \gamma}{\sum_{\{i,j\} \in \mathcal{P}} \gamma_{ij}} {}{}
\addConstraint{d_{ij}^2}{\leq \Tilde{r_b}^2 + ((100\sqrt{2})^2-\Tilde{r_b}^2)\gamma_{ij} \qquad & (i,j) \in \mathcal{E}}
\addConstraint{d_{ij}^2}{\geq (1-\gamma_{ij})4^2 \qquad & (i,j) \in \mathcal{E}}
\addConstraint{d_{ij}^2}{\leq 100^2 + 100^2\gamma_{ij} \qquad & (i,j) \notin \mathcal{E}}
\addConstraint{d_{ij}^2}{\geq (1-\gamma_{ij})(\Tilde{r_b}+\epsilon)^2 & (i,j) \notin \mathcal{E}}
\addConstraint{d_{ij}^2}{=||\overrightarrow{p}_i-\overrightarrow{p}_j||_2^2 & \{i,j\} \in \mathcal{P}}
\addConstraint{\overrightarrow{p}}{\in [-50,+50]^{n \times N}}
\addConstraint{\gamma}{\in \{0,1\}^{|\mathcal{P}|}}
\end{mini}

\subsection{QUBO instances description}

To evaluate the proposed methodology, different QUBO problem instances have been investigated. The studied problems all allow for a binary variables representation that can be mapped to the machine Hamiltonian \eqref{hamiltonian_eq}, thus, even if in the classical nomenclature of optimization problems, they are referred to with different names (\textit{i.e.}, graph coloring, protein folding problems), all problems that can be formulated as Maximum Independent Set problems \cite{tarjan1977finding} can take advantage of the proposed embedding methodology.

Starting from previous work on graph coloring problems \cite{vitaliOptimalGraphColoring2021}, solved through an iterative QUBO Maximum Independent Set (MIS) approach \cite{pasqal_optim}, the \textit{antennas dataset}, has been created: this dataset is natively UD-based, as it is built from the original positions of antennas\footnote{https://opencellid.org/} in the city of Turin and it presents an adjacency pattern set using conflict distance $D_c$: all antennas that are nearer than $D_c$ are in conflict and as a consequence introduce a quadratic penalty term in the MIS QUBO formulation. The \textit{antennas dataset} contains at most 87 antennas, and as each antenna is represented by a qubit in the MIS QUBO formulation, this implies that at most 87 qubits should be embedded. Anyway, according to the distance $D_c$, the \textit{antennas dataset} could be separated into connected components and the respective QUBO instances can be embedded separately. The \textit{antennas dataset} provides an example for QUBO problems that rely on a precise topology to obtain starting positions $(x_i,y_i),\ \forall i \in \mathcal{V}$ scaled to the quantum register domain.  Problems' instances of this kind will be hereafter named \textit{QUBO MIS antennas} problems.

Then, other not UD-based problems have been investigated. For instance, the protein folding problem aims to find the folding of a protein chain made of hydrophilic and hydrophobic amino acids. The chain naturally folds to bring as many hydrophobic acids as possible close together, this is modelled by maximizing the number of hydrophobic acid matchings, which in the optimization problem are modelled as binary variables $\delta_{ij}$. So, starting from the \textit{protein folding} formulation in \cite{williams2013model}, the equivalent QUBO formulation is proposed:
\begin{equation}
\begin{aligned}
& \underset{\delta_{ij}}{\text{min}}
& & - \sum_{i>j}  \delta_{ij} + P \sum_{ij,kh \in S} \delta_{ij}\delta_{kh}\\
& \text{s.t.}
& & \delta_{ij}\in \{0,1\}
\end{aligned}
\end{equation}
\begin{equation}
    S=\{ij,kh |\ i \leq f < j,\ f \neq \frac{i+j-1}{2},\ f = \frac{k+h-1}{2} \in \mathbb{N} \}
\end{equation}
The QUBO instances belonging to this class of problems will be referred to as \textit{QUBO protein folding} samples and the possible matchings $\delta_{ij}$ will be named after the hydrophobic acids positions.

Finally, a QUBO formulation that fits the Hamiltonian formulation of eq.~\eqref{hamiltonian_eq}, can be provided also for a one-hot-encoding modelled graph coloring problem \cite{glover2019quantum}. Problems' instances belonging to this class will be referred to as \textit{QUBO graph coloring} problems.

For both the \textit{QUBO protein folding} and the \textit{QUBO graph coloring} problems, the initial positions of the qubits $(x_i,y_i),\ \forall i \in \mathcal{V}$ are computed through the Fruchterman-Reingold method.\\

To describe the characteristics of the embeddings, the following notation is introduced
\begin{itemize}
    \item $|K_{max}|$, the size of the estimated maximum clique
    \item $r_{min}^0$, minimum distance among qubits in the initial configuration
    \item $r_{min}^f$, minimum distance among qubits in the final configuration
    \item $r_{max}^0$, maximum distance among qubits in the initial configuration
    \item $r_{max}^f$, maximum distance among qubits in the final configuration
    \item $r_{adj}^0$, maximum distance among adjacent qubits in the initial configuration
    \item $r_{adj}^f$, maximum distance among adjacent qubits in the final configuration
    \item $r_{\overline{adj}}^0$, minimum distance among non-adjacent qubits in the initial configuration
    \item $r_{\overline{adj}}^f$, minimum distance among non-adjacent qubits in the final configuration
    \item $E_c$, number of epochs used by the GEAN model to find a feasible embedding
    \item $Gurobi\_sol$, whether the \textit{Ipopt} solver is able to retrieve a feasible solution 
\end{itemize}

\subsection{QUBO instances embeddings}

\begin{figure*}[ht!] 
    \centerline{
    \includegraphics[width=1.0\linewidth]{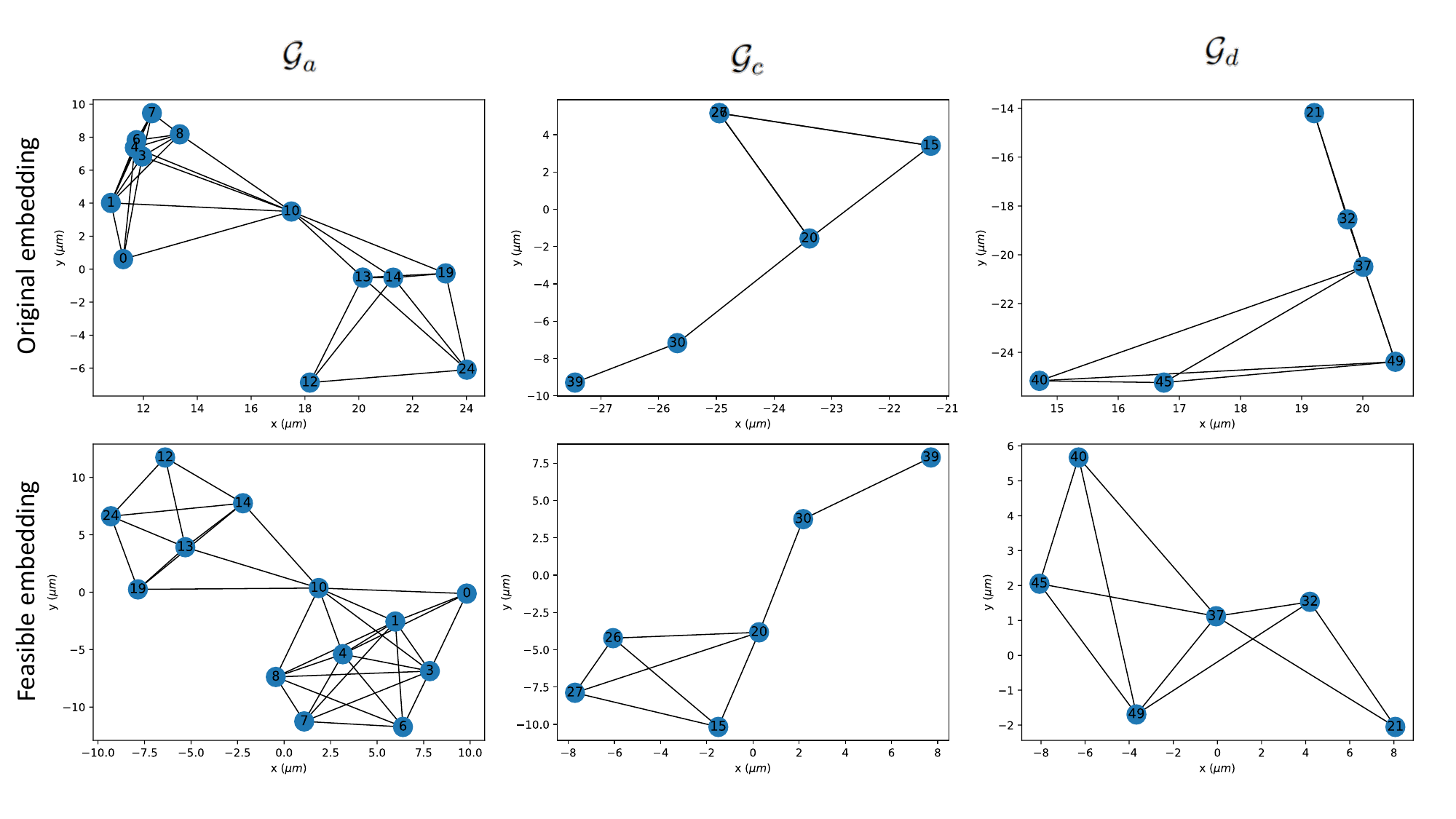}}
    \caption{Transformation into feasible 2D embeddings of the $\mathcal{G}_a$, $\mathcal{G}_c$ and $\mathcal{G}_d$ connected components for the \textit{QUBO MIS antennas} problem.}
    \label{fig:2d_antennas}
\end{figure*}

The first set of results concerns \textit{QUBO MIS antennas} problems, embedded in the 2D register. Setting a conflict distance $D_c=130\ m$, all the connected components of the corresponding MIS QUBO problems were successfully embedded onto the quantum register. In particular, $10$ out of the $87$ initial vertexes were isolated nodes, the other vertexes could be divided into 17 connected components. Among the connected components, there were 6 $K_2$, 1 $K_3$ and 1 $K_5$ graphs, whose feasible embedding can be trivially obtained and 3 graphs with $3$ vertexes that had yet a feasible embedding from the scaling operation. All other connected components required the intervention of the GEAN model to reach feasibility as summarized in Table~\ref{tab:MIS_antennas}.
From the results, it can be noticed that the unfeasible condition of the initial position is related to too near qubits that violate constraint~\eqref{eq:min}. However, applying a uniform magnification to work around the issue with the $r_{min}^0$ would just shift the issue: for instance, considering graph $\mathcal{G}_c$, $r_{min}^0$ is 3 order of magnitude below the minimum allowed distance of $4\ \mu m$, a uniform magnification would make the constraint~\eqref{eq:min} satisfied, but would violate constraints~\eqref{eq:max} and~\eqref{eq:adj}. Fig.~\ref{fig:2d_antennas} shows the embedding transformations on the hardest to embed, in terms of $E_c$, connected components of the \textit{QUBO MIS antennas} problem.

\begin{table*}[ht!]
\caption{Results obtained with the GEAN model to embed the \textit{QUBO MIS antennas} instances with a conflict distance $D_c$ set to $130\ m$.}
\begin{center}
\begin{tabular}{c|ccccccccccccc}
\toprule
\textbf{$\mathcal{G}$} & \textbf{$|\mathcal{V}|$}& \textbf{$d_{max}$} & 
\textbf{$|K_{max}|$}&
\textbf{$r_{min}^0$}&
\textbf{$r_{min}^f$}&
\textbf{$r_{max}^0$}&
\textbf{$r_{max}^f$}&
\textbf{$r_{adj}^0$}&
\textbf{$r_{adj}^f$}&
\textbf{$r_{\overline{adj}}^0$}&
\textbf{$r_{\overline{adj}}^f$} &
\textbf{$E_c$} &
\textbf{$Gurobi\_sol$} \\
& & & & ($\mu m$) & ($\mu m$) & ($\mu m$) & ($\mu m$) & ($\mu m$) & ($\mu m$) & ($\mu m$)& ($\mu m$) & &\\
\midrule
$\mathcal{G}_a$ & 13 & 8 & 5 & 0.454 & 4.006 & 19.457 &  26.725 & 7.070 & 9.984 & 7.193 & 10.634 & 733 & Yes\\
$\mathcal{G}_b$ & 7 & 4 & 4 &  2.634 & 4.004 & 15.956 & 21.265 & 6.873 & 10.090 & 7.265 & 11.155 & 71 & Yes\\
$\mathcal{G}_c$ & 6 & 4 & 4 &  0.001 & 4.013 & 14.663 & 22.046 & 6.899 & 8.943 & 8.731 & 11.462 & 384 & Yes \\
$\mathcal{G}_d$ & 6 & 5 & 3 &  1.960 & 4.028 & 11.855 & 16.650 & 7.067 & 8.724 & 7.334 & 11.273 & 374 & Yes \\
$\mathcal{G}_e$ & 4 & 3 & 3 & 1.198 & 4.037 & 7.799 & 8.547 & 6.952 & 6.511 & 7.799 & 8.547 & 98 & Yes\\
$\mathcal{G}_f$ & 5 & 4 & 4 & 0.855 & 4.061 & 7.255 & 13.473 & 6.670 & 9.865 & 7.255 & 13.473 & 261 & Yes\\
$\mathcal{G}_g$ & 7 & 5 & 4 & 2.559 & 4.015 & 13.137 & 21.496 & 6.572 & 8.616 & 7.418 & 11.139 & 281 & Yes\\
\bottomrule
\end{tabular}
\label{tab:MIS_antennas}
\end{center}
\end{table*}

Concerning the instances of \textit{QUBO protein folding} problems, feasible embeddings were reached for all the proposed instances. The graphs associated with the QUBO problems will be referred to as $\mathcal{G}_{aa\_ha}$, with $aa$ as the number of the amino acids of the entire chain and $ha$ the number of the hydrophobic ones. In particular, the results concerned the following instances:
\begin{itemize}
    \item $\mathcal{G}_{12\_6}$ with hydrophobic amino acids placed at positions 1, 2, 3, 5, 11, 12 
    \item $\mathcal{G}_{17\_7}$ with hydrophobic amino acids placed at positions 1, 2, 5, 6, 10, 12, 17
    \item $\mathcal{G}_{22\_8}$ with hydrophobic amino acids placed at positions 1, 3, 5, 6, 9, 10, 11, 17
\end{itemize}

\begin{figure*}[ht!] 
    \centerline{
    \includegraphics[width=1.0\linewidth]{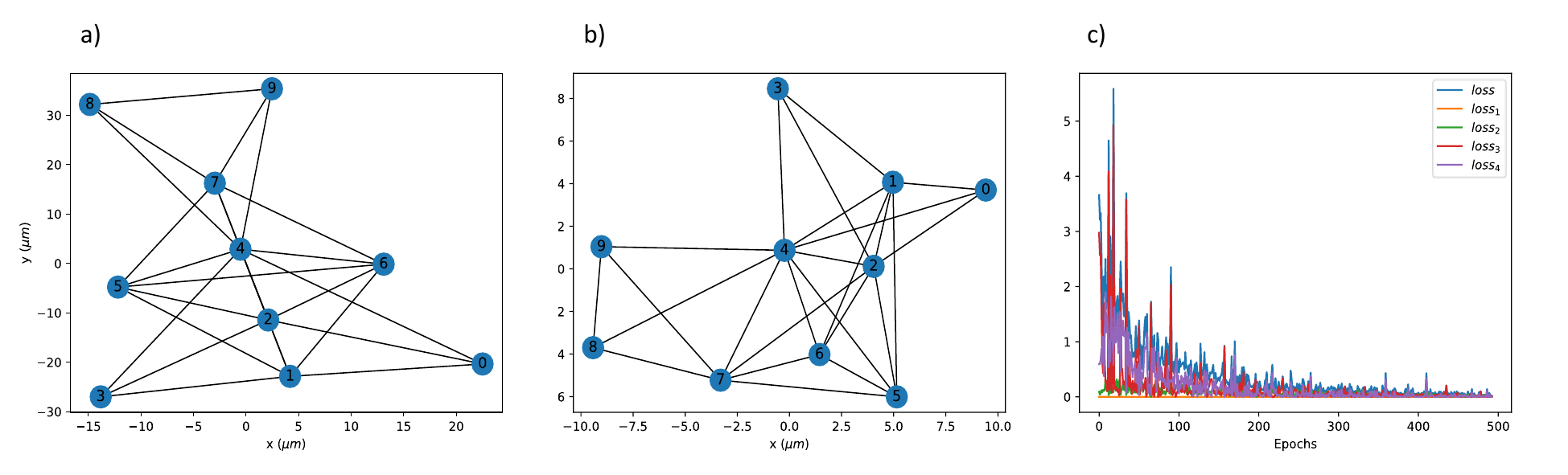}}
    \caption{Embedding for the \textit{QUBO protein folding} problem corresponding to $\mathcal{G}_{17\_7}$: the initial position, fig. \ref{fig:protein}a), are computed through Fruchterman-Reingold method and are mapped into a feasible 2D UD embedding reported in fig. \ref{fig:protein}b). Fig. \ref{fig:protein}c) shows the behaviour of the $loss$ function along the epochs.}
    \label{fig:protein}
\end{figure*}

In these cases, the initialization of the coordinates $(x_i,y_i) \ \forall i \in \mathcal{V}$ is performed through the Fruchterman-Reingold method with the equilibrium point between the attractive and the repulsive forces set to $k=4\ \mu m$. As can be noticed from Table~\ref{tab:protein2d}, the initial positions lack the unit disk property, as $r_{adj}^{0} > r_{\overline{adj}}^{0}$. Moreover, the absence of the control on the maximum distance between adjacent vertexes leads to $r_{adj}^{0}>\Tilde{r_b}$. The difficulty of the UD embedding task is reflected in the loss function behaviour, see Fig.~\ref{fig:protein}, most of the loss contribution comes from the adjacency pattern requirements. In fact, averaging on the overall $loss$ score, $37\%$ and $31\%$ contributions come respectively from $loss_3$ and $loss_4$.\\

\begin{table*}[ht!]
\caption{Results obtained with the GEAN model to embed instances of the \textit{QUBO protein folding} problem.}
\begin{center}
\begin{tabular}{c|ccccccccccccc}
\toprule
\textbf{$\mathcal{G}$} & \textbf{$|\mathcal{V}|$}& \textbf{$d_{max}$} & 
\textbf{$|K_{max}|$}&
\textbf{$r_{min}^0$}&
\textbf{$r_{min}^f$}&
\textbf{$r_{max}^0$}&
\textbf{$r_{max}^f$}&
\textbf{$r_{adj}^0$}&
\textbf{$r_{adj}^f$}&
\textbf{$r_{\overline{adj}}^0$}&
\textbf{$r_{\overline{adj}}^f$} &
\textbf{$E_c$} &
\textbf{$Gurobi\_sol$} \\
& & & & ($\mu m$) & ($\mu m$) & ($\mu m$) & ($\mu m$) & ($\mu m$) & ($\mu m$) & ($\mu m$)& ($\mu m$) & &\\
\midrule
$\mathcal{G}_{12\_6}$ & 5 & 4 & 3 & 27.344 & 4.000 & 74.611 &   11.324 & 55.524 & 9.853 & 27.344 & 11.138 & 187 & Yes\\
$\mathcal{G}_{17\_7}$ & 10 & 9 & 4 &  11.601 & 4.056 & 64.463 & 20.237 & 32.719 & 10.255 & 22.277 &10.606 & 492 & No\\
$\mathcal{G}_{22\_8}$ & 9 & 7 & 4 & 9.438 & 4.104 & 79.640 & 23.329 & 27.957 & 10.217 & 27.215 & 10.514 & 404 & Yes\\
\bottomrule
\end{tabular}
\label{tab:protein2d}
\end{center}
\end{table*}

\begin{figure*}[ht!] 
    \centerline{
    \includegraphics[width=1.0\linewidth]{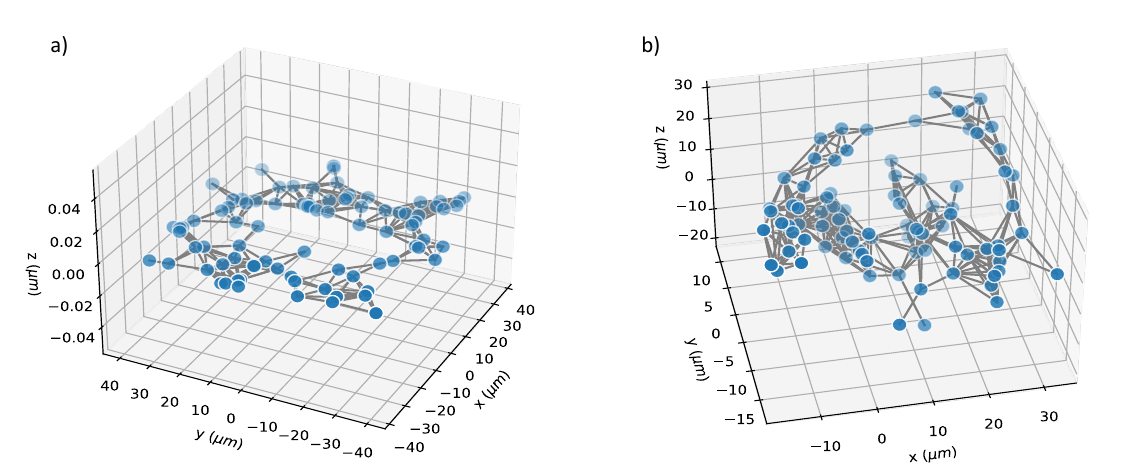}}
    \caption{Embedding for the full \textit{QUBO MIS antennas} problem in 3D: the initial positions with z-coordinates set to $0$, fig. \ref{fig:3d_antennas}a), are mapped into a feasible 3D embedding, fig. \ref{fig:3d_antennas}b).}
    \label{fig:3d_antennas}
\end{figure*}

Sometimes the limitations on the 2D embeddings do not allow for a feasible UD embedding. Hence, here are reported the instances whose adjacency pattern could not be fully respected in a 2D configuration, but that achieved feasibility through the exploitation of the third dimension.
An example comes from the \textit{QUBO MIS antennas} case with an increased conflict distance $D_c=250\ m$, which makes the graph of the corresponding QUBO model form a unique connected component, so the UD embedding considers $87$ qubits to place into a sphere domain of radius $50\ \mu m$, see Fig. \ref{fig:3d_antennas}.

The increased dimensionality allows the embedding to deal with a more complex graph that has $d_{max}=14$ and $|K_{max}|=10$. The effect of the GEAN model on the coordinates transformation has a significant impact on the feasibility of the embedding: the minimum distance among qubits is increased by $3$ orders of magnitude, from $r_{min}^0= 0.001\ \mu m$ to $r_{min}^f=4.007\ \mu m$, and the distance among adjacent qubits is reduced, from $r_{adj}^0=13.552\ \mu m$ to $r_{adj}^f=10.134\ \mu m$, within $2120$ epochs. Moreover, the UD property is still preserved, as $r_{\overline{adj}}^f$ is $10.415\ \mu m$ and the maximum distance constraint is largely satisfied as $r_{max}^f=53.300\ \mu m$. The \textit{Gurobi} solver could not provide a feasible solution within a comparable time.

\begin{figure*}[ht!] 
    \centerline{
    \includegraphics[width=1.0\linewidth]{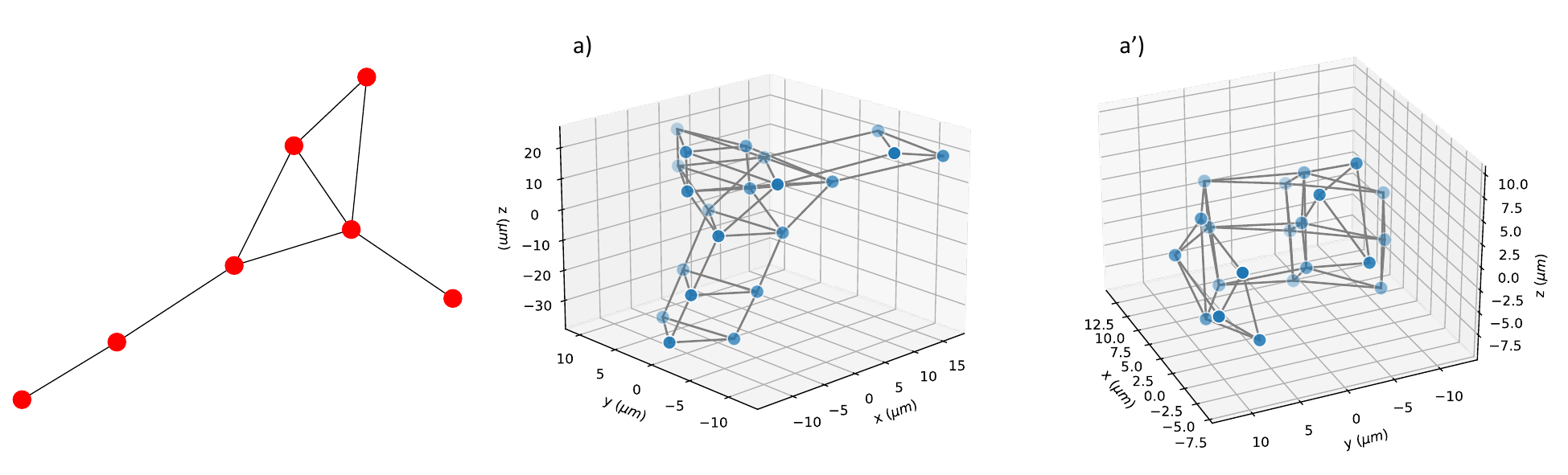}}
    \caption{\textit{QUBO graph coloring} embedding in 3D corresponding to $\mathcal{G}_{GC}$: starting from the red graph on the left, the one-hot-encoding QUBO formulation is built to take into account $3$ colors. The initial qubits positions shown on fig. \ref{fig:gc_embedding}a) are mapped into a feasible configuration as in fig. \ref{fig:gc_embedding}a').}
    \label{fig:gc_embedding}
\end{figure*}

Another example is from the \textit{QUBO graph coloring} problem: from the red graph in Fig.~\ref{fig:gc_embedding}, a three colors graph coloring QUBO problem was formulated. In this case, the graph $\mathcal{G}_{GC}$ that is associated with the QUBO formulation has $21$ vertexes, $d_{max}=6$, $|K_{max}|=3$, corresponding to the clique that is associated with the exclusivity constraint of colors assigned to a vertex. 
The effect of the GEAN model transformation can be summed up as follows:
\begin{align*}
    &r_{min}^0=8.948\ \mu m &\longrightarrow&\qquad r_{min}^f=4.183\ \mu m \\
    &r_{max}^0=67.980\ \mu m &\longrightarrow&\qquad r_{max}^f=28.545\ \mu m \\
    &r_{adj}^0=19.217\ \mu m &\longrightarrow&\qquad r_{adj}^f=10.242\ \mu m \\
    &r_{\overline{adj}}^0=9.137\ \mu m &\longrightarrow&\qquad r_{\overline{adj}}^f=10.265\ \mu m
\end{align*}
The feasible UD embedding was reached within $767$ epochs, whilst the \textit{Gurobi} solver did not reach a feasible embedding within the available wall-time.

A final observation about the power of this embedding approach concerns its running speed: since the input of each training step consists of just one sample, even many epochs can be performed in a short time. Implementing and running the model on a standard consumer machine not equipped with dedicated acceleration hardware, it was possible to obtain solutions for both the 2D and the 3D embeddings instances within $2$ minutes also for the biggest and most complex instance, \textit{i.e.}, embedding of \textit{QUBO MIS antennas} in 3D. This capability to retrieve solution in such short times represent an advantage over traditional approaches for the solution of UD graph problems, as a matter of fact, the \textit{Gurobi} solver was not able to provide a feasible embedding for instances considering $10$, $21$ and $87$ qubits.

\section{Conclusion}
 
Summing up, the GEAN model shortens the gap between the theoretical QUBO problems formulation and the actual quantum optimization, by allowing UD embedding to match with the machine’s effective Hamiltonian. The proposed model relies on a strong characterization of a specific quantum register, but it can be easily adapted to match other requirements, at different levels. Nevertheless, our solution is not constrained to a fixed lattice structure and requires the minimum number of physical qubits to provide feasible embeddings. Thus, it provides an efficient tool to test QUBO problems of non-trivial scale on quantum hardware that cannot handle many qubits. The success of the GEAN model over other classical approaches is enhanced by the tight control of the distances, relying on loss function definition, and the exploitation of very effective optimizers for neural networks' training. Another key feature of the proposed solution is the multi-dimensional readiness, the extension from the 2D to the 3D case has yet been presented here, but a general $N$-dimensions definition, targeting distances' computation as outputs, is straightforward. As a matter of fact, this methodology can pave the way to other problems' solutions, not only embedding related ones.\\

Future directions for this work will consider a further characterization of the QUBO problems that admits a feasible UD embedding, both for the 2D and the 3D cases. A more detailed comparison with state-of-the-art solvers or heuristic approaches will be also performed, thus gathering more insights on the scalability and performance of the GEAN model.
Moreover, other stopping criteria and convergence analysis along the epochs will be explored to increase the gap between $r_{adj}^f$ and $r_{\overline{adj}}^f$, with the aim of differentiating better adjacent from non-adjacent qubits in the embedding configuration.
Finally, a test phase on the real machine for some classes of problems, mainly \textit{QUBO MIS antennas}' instances, will bring further validation to the embedding methodology. The validation on the real hardware will also support a study to quantify the error tolerance for the qubits' placements, concerning  the expected Rydberg blockade effect.  
\section*{Acknowledgment}

This work has been supported by the CINECA consortium and by Pasqal in the context of the Iscra-C project QOPS, resulting from a joint collaboration.


\printbibliography 

@article{pasqal_optim,
   title={Qualifying quantum approaches for hard industrial optimization problems. A case study in the field of smart-charging of electric vehicles},
   volume={8},
   ISSN={2196-0763},
   url={http://dx.doi.org/10.1140/epjqt/s40507-021-00100-3},
   DOI={10.1140/epjqt/s40507-021-00100-3},
   number={1},
   journal={EPJ Quantum Technology},
   publisher={Springer Science and Business Media LLC},
   author={Dalyac, Constantin and Henriet, Loïc and Jeandel, Emmanuel and Lechner, Wolfgang and Perdrix, Simon and Porcheron, Marc and Veshchezerova, Margarita},
   year={2021},
   month={5} }

@article{henrietQuantumComputingNeutral2020,
  title = {Quantum Computing with Neutral Atoms},
  author = {Henriet, Lo{\"i}c and Beguin, Lucas and Signoles, Adrien and Lahaye, Thierry and Browaeys, Antoine and Reymond, Georges-Olivier and Jurczak, Christophe},
  year = {2020},
  month = sep,
  journal = {Quantum},
  volume = {4},
  pages = {327},
  issn = {2521-327X},
  doi = {10.22331/q-2020-09-21-327},
  langid = {english}
}

@article{silverioPulserOpensourcePackage2022,
  title = {Pulser: {{An}} Open-Source Package for the Design of Pulse Sequences in Programmable Neutral-Atom Arrays},
  shorttitle = {Pulser},
  author = {Silv{\'e}rio, Henrique and Grijalva, Sebasti{\'a}n and Dalyac, Constantin and Leclerc, Lucas and Karalekas, Peter J. and Shammah, Nathan and Beji, Mourad and Henry, Louis-Paul and Henriet, Lo{\"i}c},
  year = {2022},
  month = jan,
  journal = {arXiv:2104.15044 [quant-ph]},
  eprint = {2104.15044},
  eprinttype = {arxiv},
  primaryclass = {quant-ph},
  archiveprefix = {arXiv}
}

@article{glover2018tutorial,
  title={A tutorial on formulating and using QUBO models},
  author={Glover, Fred and Kochenberger, Gary and Du, Yu},
  journal={arXiv preprint arXiv:1811.11538},
  year={2018}
}

@article{lucasIsingFormulationsMany2014,
  title = {Ising Formulations of Many {{NP}} Problems},
  author = {Lucas, Andrew},
  year = {2014},
  journal = {Frontiers in Physics},
  volume = {2},
  issn = {2296-424X},
  doi = {10.3389/fphy.2014.00005}
}

@article{serretSolvingOptimizationProblems2020,
  title = {Solving Optimization Problems with {{Rydberg}} Analog Quantum Computers: {{Realistic}} Requirements for Quantum Advantage Using Noisy Simulation and Classical Benchmarks},
  shorttitle = {Solving Optimization Problems with {{Rydberg}} Analog Quantum Computers},
  author = {Serret, Michel Fabrice and Marchand, Bertrand and Ayral, Thomas},
  year = {2020},
  month = nov,
  journal = {Physical Review A},
  volume = {102},
  number = {5},
  pages = {052617},
  issn = {2469-9926, 2469-9934},
  doi = {10.1103/PhysRevA.102.052617},
  langid = {english}
}

@article{breuUnitDiskGraph1998,
  title = {Unit Disk Graph Recognition Is {{NP-hard}}},
  author = {Breu, Heinz and Kirkpatrick, David G.},
  year = {1998},
  month = jan,
  journal = {Computational Geometry},
  volume = {9},
  number = {1-2},
  pages = {3--24},
  issn = {09257721},
  doi = {10.1016/S0925-7721(97)00014-X},
  langid = {english}
}

@article{PhysRevLett.104.010502,
  title = {Entanglement of Two Individual Neutral Atoms Using Rydberg Blockade},
  author = {Wilk, T. and Ga\"etan, A. and Evellin, C. and Wolters, J. and Miroshnychenko, Y. and Grangier, P. and Browaeys, A.},
  journal = {Phys. Rev. Lett.},
  volume = {104},
  issue = {1},
  pages = {010502},
  numpages = {4},
  year = {2010},
  month = {1},
  publisher = {American Physical Society},
  doi = {10.1103/PhysRevLett.104.010502},
  url = {https://link.aps.org/doi/10.1103/PhysRevLett.104.010502}
}

@article{ciampini2015ultracold,
  title={Ultracold Rubidium Atoms Excited to Rydberg Levels},
  author={Ciampini, Donatella and Morsch, Oliver and Arimondo, Ennio},
  journal={Journal of Atomic, Molecular, Condensed Matter and Nano Physics},
  volume={2},
  number={3},
  pages={161--167},
  year={2015}
}

@article{park2017general,
  title={General heuristics for nonconvex quadratically constrained quadratic programming},
  author={Park, Jaehyun and Boyd, Stephen},
  journal={arXiv preprint arXiv:1703.07870},
  year={2017}
}

@article{fruchterman1991graph,
  title={Graph drawing by force-directed placement},
  author={Fruchterman, Thomas MJ and Reingold, Edward M},
  journal={Software: Practice and experience},
  volume={21},
  number={11},
  pages={1129--1164},
  year={1991},
  publisher={Wiley Online Library}
}

@article{doi:10.1080/10556788.2017.1350675,
author = {Sourour Elloumi and Amélie Lambert},
title = {Global solution of non-convex quadratically constrained quadratic programs},
journal = {Optimization Methods and Software},
volume = {34},
number = {1},
pages = {98-114},
year  = {2019},
publisher = {Taylor & Francis},
doi = {10.1080/10556788.2017.1350675},
URL = { 
        https://doi.org/10.1080/10556788.2017.1350675
},
eprint = { 
        https://doi.org/10.1080/10556788.2017.1350675
}}

@article{kimRydbergQuantumWires2021,
  title = {Rydberg {{Quantum Wires}} for {{Maximum Independent Set Problems}} with {{Nonplanar}} and {{High-Degree Graphs}}},
  author = {Kim, Minhyuk and Kim, Kangheun and Hwang, Jaeyong and Moon, Eun-Gook and Ahn, Jaewook},
  year = {2021},
  month = sep,
  journal = {arXiv:2109.03517 [physics, physics:quant-ph]},
  eprint = {2109.03517},
  eprinttype = {arxiv},
  primaryclass = {physics, physics:quant-ph},
  archiveprefix = {arXiv}
}

@inproceedings{kuhnUnitDiskGraph2004a,
  title = {Unit Disk Graph Approximation},
  booktitle = {Proceedings of the 2004 Joint Workshop on {{Foundations}} of Mobile Computing  - {{DIALM-POMC}} '04},
  author = {Kuhn, Fabian and Moscibroda, Thomas and Wattenhofer, Roger},
  year = {2004},
  pages = {17},
  publisher = {{ACM Press}},
  address = {{Philadelphia, PA, USA}},
  doi = {10.1145/1022630.1022634},
  isbn = {978-1-58113-921-1},
  langid = {english}
}

@article{bhoreUnitDiskRepresentations2021,
  title = {Unit {{Disk Representations}} of {{Embedded Trees}}, {{Outerplanar}} and {{Multi-Legged Graphs}}},
  author = {Bhore, Sujoy and L{\"o}ffler, Maarten and Nickel, Soeren and N{\"o}llenburg, Martin},
  year = {2021},
  month = aug,
  journal = {arXiv:2103.08416 [cs]},
  eprint = {2103.08416},
  eprinttype = {arxiv},
  primaryclass = {cs},
  archiveprefix = {arXiv}
}

@article{caiPracticalHeuristicFinding2014,
  title = {A Practical Heuristic for Finding Graph Minors},
  author = {Cai, Jun and Macready, William G. and Roy, Aidan},
  year = {2014},
  month = jun,
  journal = {arXiv:1406.2741 [quant-ph]},
  eprint = {1406.2741},
  eprinttype = {arxiv},
  primaryclass = {quant-ph},
  archiveprefix = {arXiv}
}

@article{Gui2019,
author = {Gui, Jihong and Jiang, Zhipeng and Gao, Suixiang},
doi = {10.1109/ACCESS.2018.2885313},
issn = {21693536},
journal = {IEEE Access},
pages = {203--214},
publisher = {Institute of Electrical and Electronics Engineers Inc.},
title = {{PCI Planning Based on Binary Quadratic Programming in LTE/LTE-A Networks}},
volume = {7},
year = {2019}
}

@InProceedings{Kavlak2012,
author="Kavlak, Hakan
and Ilk, Hakki",
editor="Becvar, Zdenek
and Bestak, Robert
and Kencl, Lukas",
title="PCI Planning Strategies for Long Term Evolution Networks",
booktitle="NETWORKING 2012 Workshops",
year="2012",
publisher="Springer Berlin Heidelberg",
address="Berlin, Heidelberg",
pages="151--156",
isbn="978-3-642-30039-4"
}

@techreport{hagberg2008exploring,
  title={Exploring network structure, dynamics, and function using NetworkX},
  author={Hagberg, Aric and Swart, Pieter and S Chult, Daniel},
  year={2008},
  institution={Los Alamos National Lab.(LANL), Los Alamos, NM (United States)}
}

@article{chang2010simple,
  title={A simple proof of thue's theorem on circle packing},
  author={Chang, Hai-Chau and Wang, Lih-Chung},
  journal={arXiv preprint arXiv:1009.4322},
  year={2010}
}

@article{loshchilov2017decoupled,
  title={Decoupled weight decay regularization},
  author={Loshchilov, Ilya and Hutter, Frank},
  journal={arXiv preprint arXiv:1711.05101},
  year={2017}
}

@misc{vitaliOptimalGraphColoring2021,
title = {Towards {{Optimal Graph Coloring Using Rydberg Atoms}}},
booktitle = {The {{International Conference}} for {{High Performance Computing}}, {{Networking}}, {{Storage}}, and {{Analysis}}, {{Research}} Posters},
author = {Vitali, Giacomo and Viviani, Paolo and Vercellino, Chiara and Scarabosio, Andrea and Scionti, Alberto and Terzo, Olivier and Giusto, Edoardo and Montrucchio, Bartolomeo},
year = {2021},
copyright = {All rights reserved}
}

@book{williams2013model,
  title={Model building in mathematical programming},
  author={Williams, H Paul},
  year={2013},
  publisher={John Wiley \& Sons}
}

@article{glover2019quantum,
  title={Quantum Bridge Analytics I: a tutorial on formulating and using QUBO models},
  author={Glover, Fred and Kochenberger, Gary and Du, Yu},
  journal={4OR},
  volume={17},
  number={4},
  pages={335--371},
  year={2019},
  publisher={Springer}
}

@article{lechner2015quantum,
  title={A quantum annealing architecture with all-to-all connectivity from local interactions},
  author={Lechner, Wolfgang and Hauke, Philipp and Zoller, Peter},
  journal={Science advances},
  volume={1},
  number={9},
  pages={e1500838},
  year={2015},
  publisher={American Association for the Advancement of Science}
}

@article{choi2008minor,
  title={Minor-embedding in adiabatic quantum computation: I. The parameter setting problem},
  author={Choi, Vicky},
  journal={Quantum Information Processing},
  volume={7},
  number={5},
  pages={193--209},
  year={2008},
  publisher={Springer}
}

@book{bynum2021pyomo,
title={Pyomo--optimization modeling in python},
author={Bynum, Michael L. and Hackebeil, Gabriel A. and Hart, William E. and Laird, Carl D. and Nicholson, Bethany L. and Siirola, John D. and Watson, Jean-Paul and Woodruff, David L.},
edition={Third},
volume={67},
year={2021},
publisher={Springer Science \& Business Media}
}

@article{hart2011pyomo,
title={Pyomo: modeling and solving mathematical programs in Python},
author={Hart, William E and Watson, Jean-Paul and Woodruff, David L},
journal={Mathematical Programming Computation},
volume={3},
number={3},
pages={219--260},
year={2011},
publisher={Springer}
}

@article{qiu2020programmable,
  title={Programmable quantum annealing architectures with Ising quantum wires},
  author={Qiu, Xingze and Zoller, Peter and Li, Xiaopeng},
  journal={PRX Quantum},
  volume={1},
  number={2},
  pages={020311},
  year={2020},
  publisher={APS}
}

@article{urban2009observation,
  title={Observation of Rydberg blockade between two atoms},
  author={Urban, E and Johnson, Todd A and Henage, T and Isenhower, L and Yavuz, DD and Walker, TG and Saffman, M},
  journal={Nature Physics},
  volume={5},
  number={2},
  pages={110--114},
  year={2009},
  publisher={Nature Publishing Group}
}

@article{robertson1995graph,
  title={Graph minors. XIII. The disjoint paths problem},
  author={Robertson, Neil and Seymour, Paul D},
  journal={Journal of combinatorial theory, Series B},
  volume={63},
  number={1},
  pages={65--110},
  year={1995},
  publisher={Elsevier}
}

@article{CLARK1990165,
title = {Unit disk graphs},
journal = {Discrete Mathematics},
volume = {86},
number = {1},
pages = {165-177},
year = {1990},
issn = {0012-365X},
doi = {https://doi.org/10.1016/0012-365X(90)90358-O},
url = {https://www.sciencedirect.com/science/article/pii/0012365X9090358O},
author = {Brent N. Clark and Charles J. Colbourn and David S. Johnson}
}

@inproceedings{kuhn2003ad,
  title={Ad-hoc networks beyond unit disk graphs},
  author={Kuhn, Fabian and Wattenhofer, Rogert and Zollinger, Aaron},
  booktitle={Proceedings of the 2003 joint workshop on Foundations of mobile computing},
  pages={69--78},
  year={2003}
}

@article{syslo1979characterizations,
  title={Characterizations of outerplanar graphs},
  author={Sys{\l}o, Maciej M},
  journal={Discrete Mathematics},
  volume={26},
  number={1},
  pages={47--53},
  year={1979},
  publisher={Elsevier}
}

@book{west2001introduction,
  title={Introduction to graph theory},
  author={West, Douglas Brent and others},
  volume={2},
  year={2001},
  publisher={Prentice hall Upper Saddle River}
}

@article{picken2018entanglement,
  title={Entanglement of neutral-atom qubits with long ground-Rydberg coherence times},
  author={Picken, CJ and Legaie, R and McDonnell, K and Pritchard, JD},
  journal={Quantum Science and Technology},
  volume={4},
  number={1},
  pages={015011},
  year={2018},
  publisher={IOP Publishing}
}

@article{eguchi2022ig,
  title={Ig-VAE: Generative Modeling of Protein Structure by Direct 3D Coordinate Generation},
  author={Eguchi, Raphael R and Choe, Christian A and Huang, Po-Ssu},
  journal={Biorxiv},
  pages={2020--08},
  year={2022},
  publisher={Cold Spring Harbor Laboratory}
}

@article{barredo2016atom,
  title={An atom-by-atom assembler of defect-free arbitrary two-dimensional atomic arrays},
  author={Barredo, Daniel and De L{\'e}s{\'e}leuc, Sylvain and Lienhard, Vincent and Lahaye, Thierry and Browaeys, Antoine},
  journal={Science},
  volume={354},
  number={6315},
  pages={1021--1023},
  year={2016},
  publisher={American Association for the Advancement of Science}
}

@article{tarjan1977finding,
  title={Finding a maximum independent set},
  author={Tarjan, Robert Endre and Trojanowski, Anthony E},
  journal={SIAM Journal on Computing},
  volume={6},
  number={3},
  pages={537--546},
  year={1977},
  publisher={SIAM}
}

@article{gallagher1988rydberg,
  title={Rydberg atoms},
  author={Gallagher, Thomas F},
  journal={Reports on Progress in Physics},
  volume={51},
  number={2},
  pages={143},
  year={1988},
  publisher={IOP Publishing}
}

@article{bixby2007gurobi,
  title={The gurobi optimizer},
  author={Bixby, Bob},
  journal={Transp. Re-search Part B},
  volume={41},
  number={2},
  pages={159--178},
  year={2007}
}

\end{document}